\documentclass[a4paper,twoside]{article}

\usepackage{epsfig}
\usepackage{subcaption}
\usepackage{calc}
\usepackage{amssymb}
\usepackage{amstext}
\usepackage{amsmath}
\usepackage{amsthm}
\usepackage{multicol}
\usepackage{pslatex}
\usepackage{apalike}
\usepackage{algorithm2e}
\usepackage[bottom]{footmisc}
\usepackage{xfrac}    % Works better with other fonts

\usepackage[hyphens]{url}
\usepackage{hyperref}
\usepackage{wasysym}
\usepackage{float}
\usepackage{multirow}
\usepackage{orcidlink}

\usepackage{SCITEPRESS}     % Please add other packages that you may need BEFORE the SCITEPRESS.sty package.

\begin{document}

\title{Device-Bound vs. Synced Credentials:\\ A Comparative Evaluation of Passkey Authentication\thanks{Author version of the paper accepted 
at \href{https://icissp.scitevents.org/?y=2025}{ICISSP 2025}. %The accepted version can be found at \url{http://dx.doi.org/xx.xx/xxxxx}% and its use is subject to the terms and conditions of SciTePress.
}
}

  \author{\authorname{Andre Büttner\orcidAuthor{0000-0002-0138-366X} and Nils Gruschka\orcidAuthor{0000-0001-7360-8314}}
  \affiliation{University of Oslo, Gaustadalléen 23B, 0373 Oslo, Norway}
 \email{\{andrbut,nilsgrus\}@ifi.uio.no}
 }


\abstract{With passkeys, the FIDO Alliance introduces the ability to sync FIDO2 credentials across a user’s devices through passkey providers. This aims to mitigate user concerns about losing their devices and promotes the shift toward password-less authentication. As a consequence, many major online services have adopted passkeys. However, credential syncing has also created a debate among experts about their security guarantees.
In this paper, we categorize the different access levels of passkeys to show how syncing credentials impacts their security and availability. Moreover, we use the established framework from Bonneau et al.'s Quest to Replace Passwords and apply it to different types of device-bound and synced passkeys. By this, we reveal relevant differences, particularly in their usability and security, and show that the security of synced passkeys is mainly concentrated in the passkey provider. We further provide practical recommendations for end users, passkey providers, and relying parties.}

\onecolumn \maketitle \normalsize \setcounter{footnote}{0} \vfill

\section{Introduction}\label{sec:introduction}
The primary authentication method for online services is still a password, the same as it was at the beginning of the Web. Nowadays, further authentication factors like OTP codes often accompany it, but the underlying security and usability issues, like phishing or forgotten passwords, remain unchanged. However, a recent trend towards password-less authentication can be observed. This development is driven by big-tech companies such as Apple, Google, and Microsoft. Together with many other organizations, they formed the Fast Identity Online (FIDO) Alliance which envisions a future without passwords\footnote{\url{https://fidoalliance.org/overview/} (Last accessed: 2024-12-16)}. The FIDO2 standard enables the user to register a cryptographic public key with a service and authenticate by proving the possession of the secret credential---now widely referred to as \textit{passkey}.

Initially, the standard only considered credentials stored on one device, such as a security key or a single smartphone or computer. However, due to the concern about losing the authenticator device, FIDO2 experienced a relatively low adoption by both service providers and end users. This has led to the paradigm shift towards synced credentials that can be exchanged between devices. Consequently, an online account can still be accessed from another device when one device gets broken or lost. Synced passkeys require a \textit{passkey provider}, an additional entity in the FIDO2 ecosystem that provides access to credentials across different devices, just like password managers. Most major password managers have already implemented passkey management and, therefore, also function as passkey providers. 

The decision to pivot to synced passkeys has caused quite some confusion and criticism. Since passkeys can be transmitted between devices, there is a higher risk of remote attacks. Furthermore, passkey providers may become a \textit{single point of failure} in case they get compromised. It has been shown in previous instances where password managers have been hacked, such as the severe breach of LastPass in 2022 \cite{lastpass2022hack}, that this is an actual threat. Moreover, another question is whether passkeys mitigate the risk of account lockout if an authenticator is lost. 

So far, little research has been done on the security and usability of passkeys, and even less has considered the differences between the different types of passkeys. In this paper, we aim to systematize the knowledge and concerns regarding passkeys and present the observations from our exploratory research. Our contribution can be summarized as follows:
\begin{itemize}
    \item We categorize the different access levels of passkey credentials introduced by passkey providers and highlight how these affect security and availability.
    \item We systematically compare passwords, device-bound passkeys, and synced passkeys using the framework presented in~\cite{bonneau2012quest} and show that the security of synced passkeys is mainly concentrated in the passkey providers.
\end{itemize}
The remaining sections of this paper are organized as follows. 
In Section~\ref{sec:related_work}, related work is described. Section~\ref{sec:access_levels} addresses different passkey access levels and Section \ref{sec:passkey_evaluation} entails our comparative evaluation of different passkey types. Finally, Section~\ref{sec:discussion} includes the discussion, and Section~\ref{sec:conclusion} summarizes our findings and gives an outlook for future work.

\section{Related Work}\label{sec:related_work}
Regarding security, the FIDO2 standard has been analyzed using formal methods, uncovering potential weaknesses in the underlying protocols~\cite{barbosa2021provable,guan2022formal,bindel2023fido2}. Some articles have examined specific vulnerabilities, such as the possibility of intermediary entities that can exploit FIDO2 extensions~\cite{xu2021sdd,buttner2022protecting} and attacks against QR-code-based registration of a passkey~\cite{kim2024session}. Others have discovered local attacks against FIDO2, bypassing browser or OS protection measures~\cite{yadav2023security,mahdad2024breaching}. We extend the work above by highlighting the role of passkey providers as a crucial part of the FIDO2 ecosystem.

To the best of our knowledge, academic research has not yet investigated the security of synced passkeys. However, there are contributions from the industry with suggestions on how to evaluate risks regarding passkeys. For example, there is a blog post~\cite{iozzo2023threat} where a threat model for passkeys is proposed. While it refers to relevant aspects such as server- and client-side credential stealing, it is limited and only makes suggestions for relying party implementations. Furthermore, in a talk at the \textit{Authenticate 2023} conference hosted by the FIDO Alliance, another threat model was described that considers several risks of passkeys related to credential compromise, recovery, and sharing~\cite{youtube2024passkeythreat}. 

The usability of FIDO2 authentication is another important topic that has been addressed by several researchers.
Some studies have analyzed the usability of smartphones as FIDO2 authenticators~\cite{owens2021user,wursching2023fido2}; however, at a time when they were only stored on a single device. Furthermore, user perceptions of security keys have been studied~\cite{lyastani2020is}. Moreover, usability aspects have been analyzed with a particular focus on using FIDO2 in enterprise contexts~\cite{kepkowski2023challenges}. 
Only one study we found also considered the usability and deployability aspects of synced passkeys. Here, CISOs and FIDO2 experts were interviewed regarding their thoughts on using passkeys in companies, showing remaining concerns, such as account recovery and technical issues~\cite{lassak2024WhyArent}.

In summary, our literature review has identified a lack of research on passkeys and the influence of passkey providers and their features. Since passkey providers substantially change the way FIDO2 credentials are accessed, our paper highlights important factors for the usability, deployability, and security risks of passkeys.

\section{Access Levels of Passkeys}\label{sec:access_levels}
\begin{figure*}
    \centering
    \includegraphics[width=.65\linewidth]{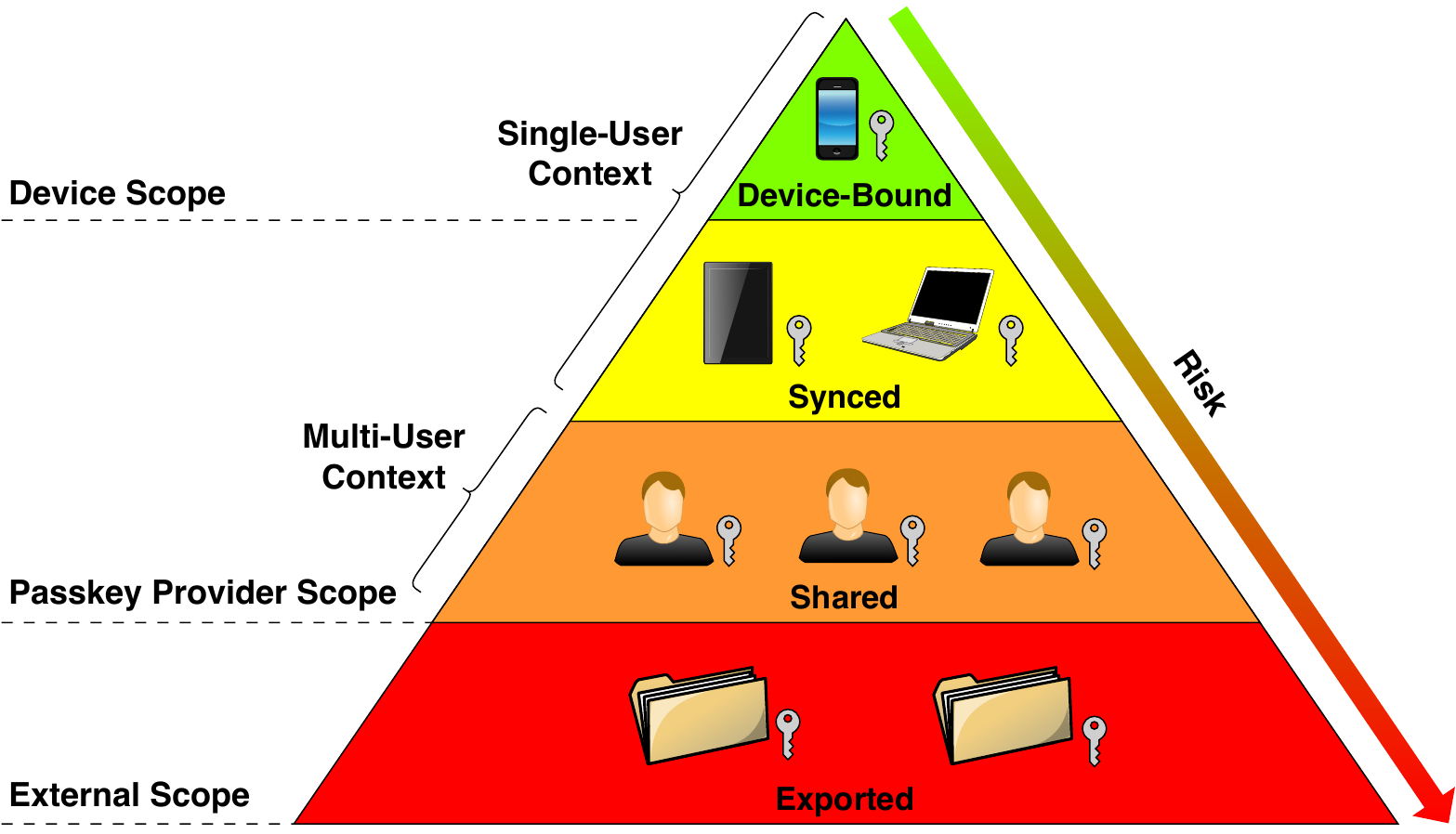}
    \caption{Access levels of passkey credentials. The estimated risk is indicated by the colors green (low-risk), yellow (low- to medium-risk), orange (medium- to high-risk ), and red (high-risk).}
    \label{fig:risk_levels}
\end{figure*}
With a passkey provider, users can distribute their credentials to different devices, share them with other users, and manually back them up. 
However, various risks can occur depending on how restricted the access to a user's passkeys is. Based on our observations, we define four different access levels of passkeys that are (1)~bound to one device, (2)~synced between the devices of a user, (3)~shared with other users, and (4)~exported in plaintext and thus accessible outside the scope of a passkey provider (see Figure~\ref{fig:risk_levels}). One can assume that a lower access restriction for a passkey leads to a larger attack surface, thus increasing the risk level. In the following, we describe each of the different access levels and their impact on passkey credentials.

\subsection{Device-Bound Passkey}
A device-bound passkey corresponds to the initial concept behind FIDO2 authentication. It is a credential that is only stored on the device where it was generated. Therefore, it does not require a passkey management tool. Since it can be protected by secure hardware, it provides the highest security guarantees, making it nearly impossible for anyone to obtain a private key. Even for an attacker with physical access to the device, it will be difficult to steal a user's device-bound passkey, provided it is protected by user verification mechanisms, e.g., a PIN or fingerprint.

\subsection{Synced Passkey}
Synced passkeys are generated on one device and distributed to others through a passkey provider. This ensures that passkey credentials remain accessible when one of the devices is damaged or lost. Since passkeys can be transmitted from one device to another, they cannot be isolated in secure hardware. This increases the chances of a remote attacker capable of installing malware on the victim's device, stealing the passkey from the memory. 

Further, passkeys must be protected when synced between the user's devices using end-to-end encryption.
Another risk is that a user's passkey provider account may get compromised. It is, therefore, crucial that a user protects their account with secure authentication methods. Thus, new devices should be authenticated to ensure an attacker cannot access the keys. Google, for instance, requires the user to enter the Google Password Manager PIN or the screen lock from an Android device that already has access to the passkeys before syncing credentials~\cite{desai2024more}. We noticed, however, that some passkey providers only require a master password, which provides access to both the passkey provider account and the database.

\subsection{Shared Passkey}
Passkey providers may also allow users to share their credentials with other users. We observed this on Apple's password manager and most third-party passkey providers. This drastically increases the attack surface of passkeys as the original owner may lose control over their credentials. As soon as a passkey is shared, one must assume this operation is irreversible. Even if a passkey provider lets a user limit or revoke access to a passkey by other users, it will not stop a malicious user from extracting the passkey out of the passkey provider. Therefore, this feature should be used carefully and only with trusted users. Particularly in enterprises, one must consider internal threats and ensure that former employees do not keep access to their passkeys.

When sharing passkeys, users also rely on protecting all user accounts that can access them. An attacker may be able to determine whether one of those accounts is only weakly protected and exploit it accordingly. Another concern could be social engineering, where a user is tricked into sharing a passkey with an account controlled by an attacker.

\subsection{Exported Passkey}
Many passkey providers allow users to export their passkeys to back them up or migrate them to another provider. 
However, as soon as passkeys leave the scope of the passkey provider, protection measures such as access control and database encryption become ineffective. Thus, the backup file should be password-encrypted, which is offered by many passkey providers, to protect its content from an attacker who gains access. This is especially relevant when storing the backup in the cloud. In any case, the export functionality should be used carefully.

The FIDO Alliance has further published a working draft for a protocol for secure credential exchange between different services. This would enable password and passkey management tools to migrate credentials with lower effort by the user and with higher protection~\cite{fido2024export}.

\section{Evaluation of Passkeys}\label{sec:passkey_evaluation}
We apply the framework presented in \cite{bonneau2012quest} to evaluate passkeys and compare different usability, deployability, and security aspects.

\subsection{Passkey Types}
Authenticators for device-bound credentials, which correspond with the original FIDO2 concept, can be divided into \textit{platform} and \textit{roaming authenticators}. A platform authenticator is embedded into the client device from which a user authenticates to a relying party. Such a device can be a computer, a tablet, or a smartphone. The operating system on the client device provides an API that enables the creation of cryptographic keys, which are generated and stored in a secure hardware module. In contrast, roaming authenticators are separate devices and physically isolated from the client device. Common examples are security keys that can be plugged into the client via USB or connected via NFC. Mobile devices can also be used as roaming authenticators in cross-device authentication contexts where the mobile device connects to the client via Bluetooth. 

Since the shift to passkeys, credentials can also be synced between different devices using a passkey provider. In this case, the FIDO Alliance distinguishes between \textit{first-} and \textit{third-party passkey providers}\footnote{\url{https://passkeys.dev/docs/reference/terms/} (Last accessed: 2024-12-16)}. First-party providers are essentially the same as built-in password managers, i.e., already provided through the operating system or browser. At the time of writing, this mainly includes Apple's and Google's respective password managers. Note that passkeys stored in Windows Hello cannot yet be synced to date. External applications supporting passkey management, such as Bitwarden, ProtonPass, or KeePassXC, are considered third-party providers. 

\subsection{Methodology}
We compare the four types of passkeys described above with each other and in addition to legacy passwords, i.e., traditional passwords memorized by the user. Since the intention behind passkeys is to replace passwords, we do not consider multi-factor authentication. Moreover, we only address the concepts and intended features of passkeys. Implementation vulnerabilities on the relying party, client, or authenticator are not considered since they can eventually affect any authentication scheme. We further assume using TLS-encrypted traffic between the client and the relying party.

The framework in \cite{bonneau2012quest} has been applied in various studies to compare different authentication methods~\cite{dressel2019securicast,lyastani2020is,arias2021inexpensive,kunke2021evaluation}. It proposes a list of 25 benefits that may (or may not) apply to particular authentication schemes, including eight benefits for usability, six for deployability, and eleven for security. 
We adopt their evaluation of legacy passwords as a baseline. For evaluating the passkeys, we discuss each of the benefits carefully to decide whether the respective passkey type \textit{offers}, \textit{partially offers}, or \textit{does not offer} it.

\subsection{Results}
\begin{table*}[t!]
\setlength{\tabcolsep}{1pt}
    \centering
   %\small
    \caption{Comparison of passwords and different passkey types based on the framework in \cite{bonneau2012quest}.}
    \label{tab:evaluation}
    \begin{tabular}{@{\hskip 1mm}l@{\hskip 3mm} l@{\hskip1mm}|@{\hskip .7mm}
    c@{\hskip .7mm}c@{\hskip .7mm}c@{\hskip .7mm}c@{\hskip .7mm}c@{\hskip .7mm}c@{\hskip .7mm}c@{\hskip .7mm}c@{\hskip .7mm} | @{\hskip .7mm}
   c@{\hskip .7mm}c@{\hskip .7mm}c@{\hskip .7mm}c@{\hskip .7mm}c@{\hskip .7mm}c@{\hskip .7mm}|@{\hskip .7mm}
    c@{\hskip .7mm}c@{\hskip .7mm}c@{\hskip .7mm}c@{\hskip .7mm}c@{\hskip .7mm}c@{\hskip .7mm}c@{\hskip .7mm}c@{\hskip .7mm}c@{\hskip .7mm}c@{\hskip .7mm}c@{\hskip .7mm}
    }
    
     & & \multicolumn{8}{c|@{\hskip .7mm}}{Usability} & \multicolumn{6}{c|@{\hskip .7mm}}{Deployability}& \multicolumn{11}{c}{Security}\\[2mm]
    & & 
    %Usability
    \rotatebox[origin=l]{90}{\textit{Memorywise-Effortless}} & \rotatebox[origin=l]{90}{\textit{Scalable-for-Users}} & \rotatebox[origin=l]{90}{\textit{Nothing-to-Carry}} & \rotatebox[origin=l]{90}{\textit{Physically-Effortless}} & \rotatebox[origin=l]{90}{\textit{Easy-to-Learn}} & \rotatebox[origin=l]{90}{\textit{Efficient-to-Use}} & \rotatebox[origin=l]{90}{\textit{Infrequent-Errors}}&  \rotatebox[origin=l]{90}{\textit{Easy-Recovery-from-Loss}} & 
    % Deployability
    \rotatebox[origin=l]{90}{\textit{Accessible}} &
    \rotatebox[origin=l]{90}{\textit{Negligible-Cost-per-User}} &
    \rotatebox[origin=l]{90}{\textit{Server-Compatible}} &
     \rotatebox[origin=l]{90}{\textit{Browser-Compatible}} &
      \rotatebox[origin=l]{90}{\textit{Mature}} &
    \rotatebox[origin=l]{90}{\textit{Non-Proprietary}} &
    % Security
    \rotatebox[origin=l]{90}{\textit{Resil.-to-Physical-Observation}} &
    \rotatebox[origin=l]{90}{\textit{Resil.-to-Targeted-Impersonation}} &
    \rotatebox[origin=l]{90}{\textit{Resil.-to-Throttled-Guessing}} &
    \rotatebox[origin=l]{90}{\textit{Resil.-to-Unthrottled-Guessing}} &
    \rotatebox[origin=l]{90}{\textit{Resil.-to-Internal-Observation}} &
    \rotatebox[origin=l]{90}{\textit{Resil.-to-Leaks-from-Other-Verifiers}} &
    \rotatebox[origin=l]{90}{\textit{Resil.-to-Phishing}} &
    \rotatebox[origin=l]{90}{\textit{Resil.-to-Theft}} &
    \rotatebox[origin=l]{90}{\textit{No-Trusted-Third-Party}} &
    \rotatebox[origin=l]{90}{\textit{Requiring-Explicit-Consent}} &
    \rotatebox[origin=l]{90}{\textit{Unlinkable}} 
    \\\hline
        \multicolumn{2}{@{\hskip .7mm}l@{\hskip .7mm}|@{\hskip .7mm}}{Legacy Password\textsuperscript{*}}& \Circle & \Circle & \CIRCLE & \Circle & \CIRCLE & \CIRCLE & \LEFTcircle &  \CIRCLE  &
        \CIRCLE & \CIRCLE & \CIRCLE & \CIRCLE & \CIRCLE & \CIRCLE & 
        \Circle & \LEFTcircle & \Circle & \Circle & \Circle & \Circle & \Circle & \CIRCLE & \CIRCLE & \CIRCLE & \CIRCLE \\\hline
        
        \multirow{2}{*}{Device-Bound} & Platform Authenticator &  \CIRCLE & \CIRCLE & \CIRCLE & \CIRCLE & \CIRCLE & \CIRCLE & \CIRCLE & \Circle  & \CIRCLE & \CIRCLE & \LEFTcircle & \CIRCLE & \CIRCLE & \CIRCLE & \CIRCLE  & \CIRCLE & \CIRCLE & \CIRCLE & \CIRCLE &  \CIRCLE& \CIRCLE & \CIRCLE &  \CIRCLE& \CIRCLE& \CIRCLE
        \\
         & Roaming Authenticator & \CIRCLE & \CIRCLE & \Circle & \Circle & \CIRCLE & \LEFTcircle & \LEFTcircle & \Circle & \CIRCLE & \LEFTcircle & \LEFTcircle & \CIRCLE & \CIRCLE & \CIRCLE & \CIRCLE & \CIRCLE & \CIRCLE & \CIRCLE & \CIRCLE & \CIRCLE & \CIRCLE & \CIRCLE &  \CIRCLE& \CIRCLE &  \CIRCLE
         \\\hline
 
  \multirow{2}{*}{Synced} & First-Party Provider & \CIRCLE & \CIRCLE &   \CIRCLE & \CIRCLE & \LEFTcircle & \CIRCLE & \CIRCLE & \LEFTcircle & \CIRCLE & \CIRCLE & \LEFTcircle & \CIRCLE & \CIRCLE & \CIRCLE & \CIRCLE & \CIRCLE & \CIRCLE & \CIRCLE & \LEFTcircle&  \CIRCLE&\CIRCLE & \CIRCLE & \Circle&\CIRCLE & \CIRCLE
 \\
         & Third-Party Provider & \CIRCLE & \CIRCLE & \CIRCLE & \CIRCLE & \Circle & \CIRCLE & \CIRCLE & \LEFTcircle & \CIRCLE & \CIRCLE & \LEFTcircle & \CIRCLE & \CIRCLE & \CIRCLE & \CIRCLE & \CIRCLE &  \CIRCLE & \CIRCLE & \LEFTcircle & \CIRCLE & \CIRCLE & \CIRCLE & \Circle & \CIRCLE & \CIRCLE
         \\\hline
    \end{tabular}\mbox{}\\\vspace{2mm}
    \CIRCLE\,= offers benefit; \LEFTcircle\,= partially offers benefit; \Circle\,= does not offer benefit\\
   * from \cite{bonneau2012quest}
\end{table*}

Table~\ref{tab:evaluation} summarizes our assessment of the different benefits of passwords and each passkey type. In the following, we describe our reasoning for the evaluation in detail.

\subsubsection{Usability}

% U1 - Memorywise-Effortless, U2 -  Scalable-for-Users
Regarding usability, all types of passkeys offer more benefits than passwords. The characteristics of being \textit{Memorywise-Effortless} and \textit{Scalable-for-Users} are inherently given because passkeys are created and stored by the authenticator, and the user does not need to remember a secret for each of the services that are accessed. This applies to all four types of passkeys. One could argue that a user must remember a PIN or master password for the authenticator devices. 
However, this has been replaced by biometric authentication for modern computers, smartphones, or security keys. Therefore, memory effort is negligible compared to remembering a password for each website. Passkeys offer scalability since registering for multiple services does not require the user to remember multiple secrets, unlike passwords.

% U3 - Nothing-to-Carry + U4 - Physically-Effortless
While passwords are \textit{Nothing-to-Carry}, they do not offer the benefit of being \textit{Physically-Effortless} since the user has to type them for every new session. A roaming authenticator is, by definition, a separate device from the client and does not offer the former benefit. Regarding the physical effort, roaming authenticators must be connected via USB, NFC, or Bluetooth. Hence, they do not provide the latter benefit either. In contrast, platform authenticators and synced passkeys offer both benefits because they are accessed on the client device from which the user wants to sign in. Moreover, they do not require physical effort because the device unlocking mechanism is used, usually a fingerprint scan or face recognition, to which the user is accustomed. 

% U5 - Easy-to-Learn 
We consider both types of device-bound passkeys to be \textit{Easy-to-Learn} since they do not require any specific knowledge by a user other than a simple interaction. %Some usability studies show that the use of FIDO2 authenticators is well-understood among test participants \textcolor{red}{[REFs!!]}. 
We assume synced passkeys with first-party providers to be less easy to learn. In some cases, the users must configure passkey syncing explicitly, for instance, when using the Google Password Manager. This is not obvious for average users, and it is also not trivial from which devices the passkeys can be accessed. Third-party providers can be even less understandable because, in addition to the challenges mentioned above that arise with first-party providers, they require an additional application or browser extension to be installed. Also, third-party providers have to be enabled explicitly on mobile devices in the device settings. When testing different providers, we observed that they provide different user interfaces, which can confuse the user, particularly when using different passkey providers simultaneously. 

% U6 - Efficient-to-Use
Once the user understands the concept, we assume passkeys to be \textit{Efficient-to-Use}. For security keys, it could be shown, e.g., in two-factor authentication contexts, that they perform significantly better than other methods~\cite{reese2019usability,reynolds2020empirical}. The only exception is roaming authenticators, such as phones, that must be connected via Bluetooth. In this case, an unavoidable delay occurs because a user first needs to scan a QR code displayed on the client device, and the roaming authenticator subsequently needs to connect to it. Therefore, we rate device-bound roaming authenticators only as partially offering this benefit. 

% U7 - Infrequent-Errors
The benefit \textit{Infrequent-Errors} is provided when using platform authenticators or synced passkeys since the user only needs to confirm the authentication procedure. There may be minor issues during biometric verification or other types of device unlocking. However, we consider this to be insignificant compared to the problems with typing passwords. Similar to the previous benefit, we argue that roaming authenticators may be error-prone when using Bluetooth devices, particularly when scanning QR codes. 

% U8 - Easy-Recovery-from-Loss
Account recovery is one of the most controversial aspects of passkeys \cite{lyastani2020is,kuchhal2023evaluating}. Although different proposals for recovery strategies for device-bound FIDO2 credentials have been evaluated \cite{frymann2020asynchronous,kunke2021evaluation} they have not been widely adopted. Thus, device-bound passkeys do not provide \textit{Easy-Recovery-from-Loss}. The FIDO Alliance has addressed this issue with synced passkeys, which are supposed to minimize the risk of account lockout. However, its effectiveness ultimately depends on how many of the user's devices store the key. Also, there is a chance that a user loses or forgets their master password. Since the database of a credential manager is usually encrypted with the master password, it might not be possible for the user to recover the data. Still, we conclude that the risk of losing access is reduced, and therefore, the benefit is partially offered for both types of synced passkeys. 

\subsubsection{Deployability}
Regarding the deployability benefits, we consider passkeys nearly as good as passwords. The first two benefits in this category \textit{Accessible} and \textit{Negligible-Cost-per-User} depend only on the client and the passkey device or provider.

% D1 - Accessible
We assume all types of passkeys to be accessible to anybody who is otherwise capable of using passwords.
% D2 - Negligible-Cost-per-User
Regarding costs, roaming authenticators such as security keys can be expensive, ranging in price from $30\$$ to over $100\$$. On the other hand, smartphones can be used as roaming authenticators and do not involve any additional costs. Thus, depending on the device type, we conclude this benefit to be partially offered. In contrast, platform authenticators and synced passkeys can be used without any cost. Some third-party providers may require passkey features to be purchased, but we found that most commercial providers offer passkey support for free.

% D3 - Server-Compatible
The remaining deployability benefits depend essentially on the FIDO2 standard and are equal for all types of passkeys. According to \cite{bonneau2012quest}, the benefit \textit{Server-Compatible} requires compatibility with text-based passwords on the server side. While passkeys do not fulfill this, we interpret this benefit rather as how expensive it is for a server to provide this authentication method. We rate this benefit partially offered due to its wide support by many services\footnote{According to 1Password, more than 200 services supported passkeys in December 2024. See \url{https://passkeys.directory/} (Last accessed: 2024-12-01)} and the availability of various WebAuthn libraries\footnote{\url{https://passkeys.dev/docs/tools-libraries/libraries/} (Last accessed: 2024-12-13)}. 
% D4 - Browser-Compatible
The benefit \textit{Browser-Compatible} is offered by passkeys since the WebAuthn standard \cite{w3c2021webauthn} is implemented by all major browsers.
% D5 - Mature
% D6 - Non-Proprietary
Because of the aforementioned broad support both on the server and client, we also consider passkeys to offer the benefit \textit{Mature}. Furthermore, passkeys are \textit{Non-Proprietary} since FIDO2 is an open standard and can be implemented by anyone.

\subsubsection{Security}
% S1 - Resilient-to-Physical-Observation
% S2 - Resilient-to-Targeted-Impersonation
% S3 - Resilient-to-Throttled-Guessing
% S4 - Resilient-to-Unthrottled-Guessing

Concerning security, all passkey types offer the first four benefits. They are \textit{Resilient-to-Physical-Observation} because observing a user authenticate with a passkey will not disclose the underlying private key used to sign the challenge. Similarly, passkeys are \textit{Resilient-to-Targeted-Impersonation} since background knowledge will not bring an attacker any closer to obtaining the passkey since it is a possession-based type of authentication. Furthermore, they are resilient to both \textit{Throttled-} and \textit{Unthrottled-Guessing} due to the underlying challenge-response protocol using modern digital signature algorithms.

% S5 - Resilient-to-Internal-Observation
For the benefit \textit{Resilient-to-Internal-Observation}, we regard the protection of the private key as highly relevant. Therefore, it is only offered entirely by device-bound authenticators. Roaming authenticators and platform authenticators isolate the private key in secure hardware and prevent potential malware on the client from stealing a key from memory. Synced passkeys cannot provide this protection entirely since the key must be accessible to passkey provider software to be shared between devices. However, since passkeys are end-to-end encrypted during transmission, they are still much more difficult to steal than intercepting passwords.

% S6 - Resilient-to-Leaks-from-Other-Verifiers
Passkeys are \textit{Resilient-to-Leaks-from-Other-Verifiers} because relying parties only store a public key and identifiers that do not correlate across services. If this information is leaked, it does not help an attacker in any way to break into a user's account on another service.

% S7 - Resilient-to-Phishing
% S8 - Resilient-to-Theft
Phishing resistance is one of the main advantages of passkeys over passwords. This is achieved through origin verification by the client and by an authenticator signing challenges only for a specific relying party. All types of passkeys implement this and consequently offer the benefit \textit{Resilient-to-Phishing}. Passkeys are also \textit{Resilient-to-Theft} as the device they are stored in is usually locked by a PIN or biometrics with a limited number of attempts.

% S9 - No-Trusted-Third-Party
Device-bound passkeys are inaccessible to third-party services since they are stored only on one device. Therefore, they offer the benefit \textit{No-Trusted-Third-Party}. This is different for synced passkeys. Although the FIDO Alliance denotes certain providers as \textit{first-party}, in terms of security, we consider all passkey providers as third-party services in relation to the user. The providers must be trusted to operate as intended and protect the user's credentials. Many aspects must be considered here, such as account protection on the passkey provider service, storage and encryption of the database, and who else is eligible to access the data. These factors can vary for each passkey provider and each user. 

% S10 - Requiring-Explicit-Consent
Passkey authentication, as with passwords, requires some user interaction and fulfills the benefit \textit{Requiring-Explicit-Consent}.
% S11 - Unlinkable
Finally, passkeys also offer the last benefit \textit{Unlinkable}. In contrast to password authentication, where users tend to reuse a password on different services, the information for passkey authentication stored by different relying parties cannot be linked to each other. The WebAuthn standard~\cite{w3c2021webauthn} even enables username-less authentication through \textit{discoverable credentials}, thus further improving the user's privacy.

\section{Discussion}\label{sec:discussion}
In this section, we reflect on our passkey evaluation, discuss limitations of our analysis, refer to the relationship between passkey providers and password managers, and offer some practical recommendations.

\subsection{Comparison of Passkey Types}
In terms of usability, roaming authenticators have several drawbacks. Platform authenticators, in contrast, offer the most usability benefits because they are integrated into the users' devices. However, both types have in common that they are not easy to recover. Since credentials are device-bound, a user cannot recover them if the authenticator device gets lost. Instead, users must configure recovery methods in their accounts, which requires much effort. As this caveat has been highlighted several times as a major obstacle to adopting FIDO2, it is highly relevant to end users. 
Synced passkeys are less likely to get lost, provided they are stored on several devices or backed up by the passkey provider or the user. Their main downside is that users might not be accustomed to them, and their functionality is less transparent than passwords. 

All types of passkeys are similar regarding their deployability since most benefits are related to the FIDO2 protocol specification. The cost of security keys is the only factor that makes roaming authenticators less favorable in comparison. As mentioned earlier, the benefit of being compatible with servers is nearly given since several libraries are available. Therefore, one can say that passkeys perform almost as well as passwords in terms of deployability.

Passwords lack many security benefits, while passkeys perform significantly better. Device-bound passkeys fulfill all of the benefits, according to our assessment. Synced passkeys provide many more benefits than legacy passwords because of asymmetric cryptography, eliminating the risk of guessing credentials and phishing. The security of synced passkeys is mainly concentrated in the passkey providers as a trusted third-party service. If they are vulnerable or act maliciously, a user's credentials may be completely unprotected, depending on how passkey credentials are accessed (as discussed in Section~\ref{sec:access_levels}). 

\subsection{Limitations}
In this paper, we present a comparison of different passkey types. We emphasize that this research is a relatively high-level qualitative analysis. The main goal is to provide an overview of the different passkey types and highlight their differences. In particular, the benefits from Section~\ref{sec:passkey_evaluation} that we evaluate as \textit{partially offered} require a more fine-grained distinction, e.g., of the respective authenticator device or a specific passkey provider. The usability aspects should be reviewed in actual usability studies to empirically assess the users' perception of the different passkey types. Furthermore, the security analysis does not cover the implementation details of passkey providers, like their authentication methods or database storage. However, our evaluation shows that many security problems with passwords are irrelevant when using passkeys and that we need to focus on specific security aspects of passkey providers.

\subsection{Passkey Provider vs. Password Manager}
Most password managers now also support passkeys and treat both passkeys and passwords the same. Consequently, many aspects of passkeys mentioned in this paper apply analogously to passwords stored in password managers. Both types of credentials can be synced across devices, shared with others, and exported. In the study from \cite{lassak2024WhyArent}, some participants even mentioned that passkey providers are essentially better password managers. In both cases, they interact with a credential management tool instead of entering their credentials. However, a common risk is that password and passkey providers can become a single point of failure, potentially compromising several user accounts through a single breach.

Nonetheless, we have focused mainly on the difference between device-bound and synced passkeys. We argue that it is essential to distinguish between passwords and passkeys because their core concepts have different security properties. Passwords can not be protected through secure hardware like device-bound passkeys. Moreover, passwords stored in a password manager are often manually created by the user and can thus be guessed by an attacker~\cite{pearman2019why}. However, the most important difference is that passwords will always be vulnerable to phishing, even when generated randomly. Users must always be cautious when entering their passwords or copy-pasting them from their password manager into a website.

\subsection{Recommendations}
For end users, it is important that passkey credentials are protected and made available. Given the security benefits, we recommend using passkeys on online services instead of passwords whenever possible. However, passkey providers should be chosen mindfully based on device compatibility to ensure credentials are synced on multiple devices. Furthermore, users should consider a provider offering strong authentication methods to ensure attackers cannot access their database. They should choose a strong master password to prevent any credential leaks even in case of an account compromise or a breach into the passkey provider. Credential sharing should be used only when necessary. Ideally, as soon as the other users do not need to access the respective account anymore, the passkey should be replaced by a new one to prevent misuse. If users manually back up their credentials by exporting them, they should store them only offline or encrypt them.

Passkey providers should offer robust access control measures to prevent unintended database access. We observed that some credential managers use the same password both for authenticating to the credential manager account and for encrypting the database. If this password is leaked, e.g., by a phishing attack, the adversary gains full access to the user's credentials. Consequently, service providers should, at best, make the users create a different password for the credential manager account or encourage multi-factor authentication. In our evaluation, we also noticed the diverse user interfaces of the passkey providers when authenticating to a relying party with a passkey. We presume that this may confuse users and that there is a need for improvements on this end. 

Relying parties do not have much influence on the users' interaction with passkey providers. However, they can decide whether they allow their users to authenticate with device-bound or synced passkeys through device attestation. If device-bound credentials are used, they should support the user in setting up proper recovery methods. We also find that passkeys are often offered as an alternative authentication method to the password rather than a replacement. Relying parties should not wrongly convey a sense of increased security. Instead, they should still recommend that their users set up a secondary authentication factor or apply risk-based authentication so the password does not undermine account security despite setting up a passkey.

\section{Conclusion}\label{sec:conclusion}
The FIDO Alliance's proposal to sync passkeys across different devices has increased the adoption of passkeys by online services but also received criticism from security experts for its potential risks.

To address this, we derived different access levels and their impact on the attack surface of passkeys and conducted a comparative evaluation of different usability, deployability, and security benefits for device-bound and synced passkeys. We argue that synced passkeys mitigate credential loss; however, only with limited guarantees as it depends on whether passkey syncing is configured on several devices or whether the passkey provider backs them up. Moreover, we show that all passkey types provide more security benefits than passwords.

Our findings support claims that synced passkeys are less secure than device-bound ones. However, the range between secure and insecure passkeys varies widely depending on their implementation and usage. Since many users are unwilling to adopt device-bound passkeys, %on a large scale, 
more effort should be made to help end users use synced passkeys most advantageously. Thus, we emphasize the need for strong authentication for passkey provider accounts, cautious use of credential-sharing, and secure storage of backups.

For future work, we propose analyzing the user's perception of synced passkeys and whether this actually increases their motivation to adopt them. Beyond this, implementation vulnerabilities of passkey provider applications should be analyzed to evaluate the protection of synced credentials in more detail.

\bibliographystyle{apalike}
{\small
\bibliography{references}}

\end{document}